\documentstyle[psfig,aaspp4]{article}

\begin{document}

\title{Spectropolarimetry of 3CR\,68.1: A Highly Inclined Quasar\altaffilmark{1}}

\author{M. S. Brotherton}
\affil{Institute of Geophysics and Planetary Physics, Lawrence Livermore National Laboratory, 7000 East Avenue, P.O. Box 808, L413, Livermore, CA 94550; mbrother@igpp.llnl.gov}
\author{Beverley J. Wills}
\affil{McDonald Observatory and Astronomy Department, University of Texas,
Austin, TX 78712; bev@astro.as.utexas.edu}
\author{Arjun Dey\altaffilmark{2}}
\affil{KPNO/NOAO\altaffilmark{3}, 950 N. Cherry Avenue, P.O. Box 26732, Tuscon, AZ 85726}
\author{Wil van Breugel}
\affil{Institute of Geophysics and Planetary Physics, Lawrence Livermore National Laboratory, 7000 East Avenue, P.O. Box 808, L413, Livermore, CA 94550; wil@igpp.llnl.gov}
\author{Robert Antonucci}
\affil{Physics Department, University of California at Santa Barbara, Santa Barbara, CA 93106; ski@ginger.physics.ucsb.edu}

\altaffiltext{1}{Based on observations at the W. M. Keck Observatory.}
\altaffiltext{2}{Currently at Johns Hopkins University, dey@skysrv.pha.jhu.edu.}
\altaffiltext{3}{The National Optical Astronomy Observatories are operated by
the Association of Universities for Research in Astronomy under cooperative
agreement with the National Science Foundation.}

\begin{abstract}

We present Keck spectropolarimetry of the highly polarized radio-loud quasar 
3CR\,68.1 ($z=1.228, V=19$).
The polarization increases from 5\% in the red (4000\AA\ rest-frame) to 
$>$10\% in the blue (1900\AA\ rest-frame).
%The narrow lines are probably not significantly polarized, but 
The broad emission lines are polarized the {\em same} as the continuum, 
which shows that 3CR\,68.1
is not a blazar as it has sometimes been regarded in the past.
We also present measurements of the emission lines and a strong, blueshifted,
associated absorption line system, as well as a detection at the emission-line
redshift of \ion{Ca}{2} K absorption, presumably from stars in the host galaxy.

3CR\,68.1 belongs to an observationally rare class of highly polarized
quasars that are neither blazars nor partially-obscured radio-quiet QSOs.
Taking into account 3CR\,68.1's other unusual properties, such as its 
extremely red spectral energy distribution and its extreme lobe dominance, we
explain our spectropolarimetric results in terms of unified models.
We argue that we have a dusty, highly inclined view of 3CR\,68.1, with 
reddened scattered (polarized) quasar light diluted by even more dust-reddened
quasar light reaching us directly from the nucleus.

\end{abstract}

\keywords{line: profiles --- quasars: emission lines --- quasars: general ---
quasars: absorption lines --- galaxies: individual: 3CR\,68.1 --- 
techniques:polarimetric} 
\vfil
\eject
\section{Introduction}

3CR\,68.1 ($z = 1.228, V = 19$) is a quasar with a unique combination
of extreme properties.  The optical-UV continuum is among the reddest of any
known quasar, with $F_{\nu} \propto \nu^{-6}$\ in the UV
(Boksenberg, Carswell, \& Oke 1976; Smith \& Spinrad 
1980). 3CR\,68.1 is one of the most highly polarized quasars, with a 
polarization in a broad bandpass from 3000-8000\AA\ of 7.5\% $\pm$ 1.3\% at 
52$\arcdeg$ $\pm$ 5$\arcdeg$ (Moore \& Stockman 1981, 1984).
The lobes of 3CR\,68.1 mark the quasar as a powerful radio source, but
the extremely faint radio core ($\approx 1$~mJy at 5\,GHz,
Bridle et al. 1994) makes this object one of the most lobe-dominant 
quasars.  Under Unified Schemes, it is the most highly inclined 3CR quasar known
(e.g., Orr \& Browne 1982; Hoekstra, Barthel \& Hes 1997).  
Extreme objects like 3CR\,68.1 may provide clues as to how their properties
may be related in less extreme AGNs.

3CR\,68.1 has been classified as a blazar because of its high polarization and
red color (e.g.,  Ledden \& Odell 1985), but we thought this was unlikely
because the degree of polarization and
position angle did not change between measurements made 6 months apart 
(Moore \& Stockman 1981, 1984), and the optical continuum level is not 
violently variable (Smith \& Spinrad 1980).  Also, it does not have the 
core-dominant radio structure always characteristic of radio-selected blazars.
In fact, in the revised 3C sample (Laing, Riley, \& Longair 1983; 
Hoekstra et al. 1997), it is the most lobe-dominant quasar with the largest 
projected lobe separation, large even among radio galaxies, 
suggesting a radio axis close to the sky plane.

Most highly polarized QSOs (HPQs) can be classified as radio-quiet
(including many BAL QSOs), or as radio-loud blazars.
The polarization mechanism in radio-quiet QSOs is scattering (e.g.,  BAL QSOs:
Hines \& Wills 1995; Goodrich \& Miller 1995; Cohen et al. 1995; Miller 1997;
Ogle 1997; Schmidt et al. 1997; Brotherton et al. 1997; and IRAS non-BAL QSOs:
Wills \& Hines 1997; Wills et al. 1992), 
perhaps with a geometry similar to that found in 
polarized Seyfert 2 galaxies like NGC 1068 (e.g.,  Miller \& Antonucci 1985,
and the review by Antonucci 1993).  In addition, many highly
polarized, low-ionization BAL QSOs and IRAS QSOs appear reddened compared with
UV-selected QSOs (Sprayberry \& Foltz 1992).

Do non-blazar radio-loud HPQs exist?  OI 287 appears to be one --
a lobe-dominant radio-loud quasar, 
optically quiescent with a constant optical
polarization of $\approx$8\% aligned parallel to the extended radio
structure (Antonucci \&
Ulvestad 1988; Rudy \& Schmidt 1988).  The broad emission lines and continuum
show similar degrees of polarization
but the narrow emission lines are unpolarized
(Goodrich \& Miller 1988).  Apparently
OI 287 is a normal quasar seen only in scattered light because a thin,
dusty disk obscures the central regions (Rudy \& Schmidt 1988;
Goodrich \& Miller 1988; Antonucci, Kinney, \& Hurt 1993).
Certainly there are buried quasars in which broad lines and continuum
are revealed only in scattered (polarized) light (e.g.,  Cygnus A, 3C 324,
3C 265, IRAS 09104+4109).  Some of these are called
radio galaxies.

We hypothesized that the polarization mechanism in 3CR\,68.1 is not optical
synchrotron as in blazars, but rather scattering with dust absorption
reddening the UV-optical spectrum.
We therefore obtained spectropolarimetry of 3CR\,68.1 with Keck II in order to 
test this hypothesis and determine the polarization mechanism.

Furthermore, as part of the complete 3CR sample, whose selection is 
predominantly based on isotropic extended radio emission, an understanding
of 3CR\,68.1 can address the question of the prevalence of red quasars 
that might escape detection by usual optical 
selection criteria (e.g.,  Webster et al. 1995).

\S\ 2 describes our observations and data reduction procedure.  
\S\ 3 presents our results and shows that
the polarization increases dramatically to shorter wavelengths, that the 
broad emission lines are polarized at the same level as the continuum.
We also discuss the absorption
spectrum and our spectroscopic detection of
the host galaxy of 3CR\,68.1.  \S\ 4 argues that scattering by 
dust or electrons is indeed the polarization mechanism, and that both the 
direct and scattered light are reddened.  We show several
models for the spectrum of 3CR\,68.1, exploring the ranges in
reddening, scattered light fraction, and host galaxy contribution.  We
briefly discuss how 3CR\,68.1 fits into unification schemes.  
\S\ 5 is a summary.

\section{Observations and Data}

On 1996 December 10 (UT), 
we observed 3CR\,68.1 with the Low Resolution Imaging Spectrometer 
(Oke et al. 1995) in spectropolarimetry mode (Good\-rich, Cohen, \& Putney 1995;
Cohen et al. 1997) on the 10 m Keck II telescope.  
The 300 line mm$^{-1}$ grating blazed at 5000\AA\ with a 1$\arcsec$ slit 
(at 226$\arcdeg$, approximately the parallactic angle) gave an effective 
resolution $\leq$ 10\AA\ (FWHM of comparison lamp lines); 
the dispersion was 2.5\AA\ pixel$^{-1}$.
The seeing just prior to the observation was 1{\farcs}1.  
The observation was broken into four 15 minute exposures, one for each
waveplate position (0$\arcdeg$, 45$\arcdeg$, 22.5$\arcdeg$, 67.5$\arcdeg$).
We observed a bright star through
UV and IR polaroid filters to measure and calibrate the 
wavelength-dependent polarization position angle changes imparted by
the instrument optics.
The polarization measurement efficiency was essentially 1.0 at all
wavelengths.  We observed the flux standard Feige 110 (at the 
parallactic angle, seeing 1.3$\arcsec$) both in direct light and with an OG 570
filter.  No order-blocking filter was used for the 3CR\,68.1 observation
because of its weak blue flux and the decreased second-order efficiency of
the grating.  Then we observed the polarized standard star
HD245310 (at the parallactic angle; Schmidt, Elston, \& Lupie 1992) in all four
waveplate positions in order to determine the zero point of the
polarization position angle.

We reduced our data to one-dimensional spectra using standard techniques within
the IRAF NOAO package. The rms uncertanties in the dispersion solution were 
0.2\AA, and we used sky lines to ensure that our zero point was accurate to 
0.1\AA.  Wavelengths are air wavelengths.  We followed standard procedures 
(Miller, Robinson, \& Goodrich 1988; Cohen et al. 1997) for calculating 
Stokes parameters and uncertanties.  We used the measurements of
Massey et al. (1988) and Massey \& Gronwall (1990) for flux calibration.

The Galactic absorption in the direction of 3CR\,68.1 is small, 
$A_V = 0.15$ (Burstein \& Heiles 1982).
We give observed quantities below, but where noted
we have corrected our data for the Galactic extinction using the 
curve of Cardelli et al. (1989).

\section{Results}

Figure 1 shows our main results, including the total flux-density spectrum 
(F$_{\lambda}$), the percentage polarization (P), the electric vector 
polarization position angle ($\theta$), and the polarized flux-density 
(${{\rm P}\over 100} \times$ F$_{\lambda}$).  
Except for the total flux spectrum, the data are binned (100\AA\ in
the continuum, but more finely around the emission lines as described below).
The bins were first formed using the one-dimensional raw-count
spectra from each waveplate position; the error bars assume Poisson noise from
the object and background sky counts, plus the CCD read noise.   These binned
data then determine the binned linear Stokes parameters, including P and
$\theta$, with errors propagated through the polarization calculation.

Figure 2 shows details of several spectral regions in total light,
including \ion{C}{3}] 
$\lambda$1909, \ion{Mg}{2} $\lambda$2800, and the narrow lines and
\ion{Ca}{2} K absorption feature at the red end of the spectrum.

\subsection{Polarimetry}

The continuum polarization increases toward shorter wavelengths, 
from $\approx$5\% at the red end of our spectrum to $>$10\% at the blue end.
Within the uncertanties $\theta$\ is constant across the entire range,
with an average $\approx$53.3$\arcdeg$\,$\pm$ 0.2$\arcdeg$.
The high polarization and position angle are consistent with the broadband
results of Moore \& Stockman (1984), suggesting that the polarization has not
varied over more than a decade.

Open squares in Fig. 1 designate finer bins coincident with some of the
more prominent spectral features (observed wavelength range of the bins):
\ion{C}{3}] $\lambda$1909 (4140-4220\AA\ and 4260-4360\AA\ for the broad 
wings, 4220-4260\AA\ for the narrower core), [\ion{Ne}{4}] $\lambda$2424 
(5385-5415\AA), \ion{Mg}{2} $\lambda$2800 (6040-6170\AA\ and 6270-6440\AA\ 
for the broad wings, 6170-6200\AA\ and 6240-6270\AA\ for the narrower core
emission, and 6200-6240\AA\ for the absorption doublet), [\ion{O}{2}]
$\lambda$3727 (8290-8325\AA), and [\ion{Ne}{3}] $\lambda$3869 (8605-8635\AA).
The \ion{Mg}{2} and \ion{C}{3}] lines appear polarized 
at similar levels and position angles as the continuum.
The strong forbidden lines, [\ion{O}{2}] $\lambda$3727 and
[\ion{Ne}{3}] $\lambda$3869 appear unpolarized in the polarized-flux spectrum,
although the uncertanties are too large to be sure.

\subsection{Emission-Line Spectrum} 

Table 1 gives parameters for the
emission lines in the total flux spectrum of 3CR\,68.1.  The uncertanties
are based on choosing extreme, but still reasonable,
high and low continua, thus representing the range in probable values.  

In both \ion{C}{3}] $\lambda$1909 and \ion{Mg}{2}\,$\lambda$2800 there are clear
inflections suggesting ``narrow'' and ``broad'' components.  In addition
to total line measurements, we fitted Gaussians to these separate components
(making a simple linear interpolation across the absorption in the \ion{Mg}{2}
profile--certainly an underestimate for the associated \ion{Mg}{2} absorber).
Again, uncertanties represent the range in measurements for reasonable continuum
choices.

We derive $z_{em}$ = 1.228, which is in agreement with that from the broad 
lines, by averaging values for several narrow lines.  This is the
same value as obtained by Smith \& Spinrad (1980).  

\subsection{Absorption-Line Spectrum} 

Table 2 gives measurements of the wavelengths and equivalent widths
for absorption lines in the total flux spectrum of 3CR\,68.1.
We identify a number of low-ionization associated absorption lines at $z_{abs} 
= 1.2258$, and a \ion{Ca}{2} K feature at $z_{abs} = 1.2287$
(\ion{Ca}{2} H is coincident with the [\ion{Ne}{3}] $\lambda$3967 line).
The measurements of the \ion{Mg}{2} absorption are more uncertain than suggested
by the table because of the narrow Mg II emission component.
The absorption lines are labeled in the upper panel of Figure 1.

Our results differ from those of Aldcroft et al. (1994) in several respects.
Their spectrum is of higher resolution (350 km s$^{-1}$), but lower 
signal-to-noise ratio.  We find a lower blueshift for the \ion{Mg}{2} doublet 
($\approx$300 km s$^{-1}$ instead of 2000 km s$^{-1}$), which seems robust in 
light of our low uncertanties and good agreement among lines, 
both in emission and absorption
(the \ion{Fe}{2} absorbers, too weak for Aldcroft et al. to have found, have
redshifts that match that of \ion{Mg}{2} and \ion{Mg}{1}).
The absolute agreement in wavelength between our results and Aldroft et al.
is good (about 100 km s$^{-1}$); the cause of the blueshift difference 
is our adopted $z_{em}$;  we use $z_{em}$ = 1.228, while Aldcroft et al. use the
less accurate $z_{em}$ = 1.240.  We do not see the intervening \ion{Mg}{2}
absorber at $z$ = 0.7750 that they report. We do not separate the
absorption feature at 7093.5\AA\ into a doublet even though our resolution
should be sufficient to do so, therefore we do not support
Aldcroft et al.'s identification of \ion{Na}{1} at $z = 0.2021$.  
We tentatively identify the line with \ion{He}{1} $\lambda$3187,
although this line is not commonly seen in absorption.

Relative to the \ion{Mg}{2} absorption, the \ion{Ca}{2} K feature is strong
and seems unlikely to be entirely interstellar.  Its redshift is larger than
that of the other absorption lines, and is consistent with $z_{em}$ for the
narrow emission lines, which suggests an origin in the stars of the host 
galaxy of 3CR\,68.1.  Comparing the equivalent width of this
feature with the feature in the elliptical galaxy template of Kinney et
al. (1996) would suggest that at 3900\AA\ rest-frame, one quarter of the total
flux in 3CR\,68.1 is contributed by the host galaxy (see also \S\ 4).

\section{Discussion}

\subsection{The Spectral Energy Distribution}

3CR\,68.1 would not be found in an UV-excess survey.  Some 10\% of quasars
in the 3CR sample, 3CR\,68.1 being the most extreme, are significantly redder
than bright optically selected quasars (Smith \& Spinrad 1980).
3CR\,68.1 has been observed in many wavebands.
Figure 3 shows the spectral energy distribution (SED) of 3CR\,68.1,
from the far IR through the 2 keV X-rays, using measurements from
the literature as noted in the caption.  3C 110, a normal `blue' lobe-dominated 
quasar, is also shown for comparison.

The UV continuum is so weak that Bregman et al. (1985) argue that the large
ratio of low-ionization to high-ionization line intensities is the result of
photoionization by a separate soft X-ray component, giving rise to an extended
partially ionized zone where strong low-ionization lines can be produced.
A simpler explanation is that the quasar continuum and line spectrum are
highly reddened, and emission-line clouds see a less obscured ionizing 
continuum than does an observer. 
The lack of a 2200\AA\ dust absorption feature is not
a strong argument against reddening:  no 2200\AA\ feature is seen in the
Small Magellanic Cloud (SMC) reddening `law' (Prevot et al. 1984);
even for Galactic dust, the 2200\AA\ feature
can be filled in by scattering, in geometries possible for 3CR\,68.1
(e.g., Witt, Thronson, \& Capuano 1992).
A 2200\AA\ dust feature is also absent in the spectra of BAL QSOs
(Sprayberry \& Foltz 1992; Hines \& Wills 1995).
Also, the composition of extragalactic dust may differ from that
found in the Milky Way.

Figure 4 compares our total flux spectrum with the photometry
of Neugebauer et al. (1979) and Smith \& Spinrad (1980).  There is
good agreement at blue wavelengths, and a ($\approx$20\%)
difference at red wavelengths.  Given the uncertanties in the flux 
calibration and our narrow slit, the general agreement is satisfactory.
It seems probable that 3CR\,68.1 has not varied significantly during the last
twenty years.

\subsection{The Polarization Mechanism}

The polarization mechanism in 3CR\,68.1 is probably
scattering by a non-spherical distribution of either small dust grains
or electrons. 
Other mechanisms can be ruled out:
\begin{enumerate}
\item{Polarized synchrotron radiation: This mechanism fails to
account for the broad-emission lines polarized the same as the continuum.}
\item{Galactic interstellar polarization: The interstellar polarization toward
two stars close to the line of sight, SAO 55659 and SAO 55667 ($<$1$\arcdeg$
from 3CR\,68.1), is $\leq$0.4\% (D. Wills, private communication).
}
\item{Host Galaxy interstellar polarization: The observed 
polarization shape and maximum differ significantly from a Serkowski Law, and
as we show below, the maximum polarization is significantly greater than 
9$\times E(B-V)$, the polarization observed toward aligned dust-reddened lines
of sight in the Milky Way.}

\end{enumerate}

Scattering by electrons or small dust grains (2$\pi a/\lambda$ $<< 1$)
can result in  very high polarization (\S\ 4.3) that is essentially 
wavelength independent.  Large grains, which polarize light
with a stronger wavelength dependence, scatter less efficiently.

We propose that the polarized spectrum arises from scattering of central quasar
light, and that
the rise in polarization to shorter wavelengths results from dilution
by a redder spectrum.  We further propose that the diluting spectrum is 
the quasar spectrum seen directly but reddened by dust,
with some possible contamination by
starlight from the host galaxy.  It is likely that interstellar
polarization from our galaxy or the host galaxy is small, so we do not
include it.
Hence the total observed spectrum of 3CR\,68.1 can be described:
\begin{equation}
Total_{\lambda} = S_{\lambda,A_S} + D_{\lambda,A_D} + H_{\lambda} + NLR,
\end{equation}
\noindent
where $S_{\lambda,A_S}$\ is a scattered QSO spectrum reddened by an extinction
$A_S$, $D_{\lambda,A_D}$\ is the direct QSO spectrum reddened by an extinction
$A_D$, $H_{\lambda}$\ is the host galaxy spectrum, and $NLR$\ is the
spectrum of the narrow-line region which is unpolarized and is emitted
probably from an extended region outside or coincident with the scattering 
region.  The scattered spectrum is equal to the polarized spectrum divided 
by the polarization $p_s$, of the scattered light alone.

To illustrate quantitatively the plausibility of our picture, and to 
constrain the fraction of scattered light and reddening towards the 
central quasar, we have matched composite model spectra to the total-light
SED of 3CR\,68.1.  We first adjusted our spectrum to the photometry of
Neugebauer et al. (1979).  Free parameters are the intrinsic polarization 
of the scattered light, $p_s$, assumed to be wavelength-independent,
and the fraction of host galaxy starlight.
The strength of the \ion{Ca}{2} K absorption feature suggests up to
one quarter of the total light at 3900\AA\ rest-frame may be host
galaxy starlight.  We note that Palomar 5m image-tube 
plates show no extensions characteristic of a luminous host galaxy 
(Longair \& Gunn 1975).
We use the elliptical galaxy template of Kinney et al. (1996) to 
represent the starlight contribution.
We represent the spectrum of the unreddened central quasar by a
mean spectrum of lobe-dominated quasars (based on spectra
of the HST sample described by Wills et al. 1995).
The scattered light spectrum is derived from this mean spectrum by scaling 
and reddening it using an SMC-type extinction law
(Prevot et al. 1984) with $A_V = 0.7$\ to match the observed polarized
flux density spectrum, then dividing by $p_s$.
After subtracting the host galaxy starlight and scattered light from the SED,
we are left with a spectrum that we fit with a reddened central quasar spectrum.

Figure 5 presents four model composite spectra. Table 3 summarizes the model
parameters, along with those of two intermediate cases.
Models A, B, and C include no contribution from the 
host galaxy, and have $p_s = 80$\%, 50\%, and 20\%.
The direct line-of-sight reddening required is $A_V$ = 1.2, 1.3, and 1.6,
respectively.  Models A$_G$, B$_G$, and C$_G$ cover the same range in 
$p_s$, and include a host galaxy contribution of 25\% of the total flux 
density at 3900\AA.  The deduced line-of-sight reddening is then $A_V$ = 1.1,
1.2, and 1.5, respectively.  Also tabulated is the ratio of scattered to
direct light after correction for reddening.
Our comparison in Figure 5 extends over
the range in our templates, from rest-frame $B$ to $H$; the fits over
the range of our Keck data are in good agreement for the continuum,
but if the SED from \S\ 4.1 is correct in the near-IR, Figure 5 shows that 
models without a large elliptical host galaxy contribution are preferred. 

Assuming H$_o$ = 50 km s$^{-1}$ Mpc$^{-1}$ and q$_o$ = 0, and using the
observed spectrum and the known Galactic reddening, 
the absolute rest-frame $B$\ magnitude for 3CR\,68.1 is M($B$) = $-$26.6.
Under the parameters of model A and dereddening the total light
spectrum by $A_V = 1.2$, the intrinsic unreddened M($B$) = $-$28.2.
This result is not very model dependent.
Therefore 3CR\,68.1 is one of the most optically luminous 3CR quasars,
more in line with its radio luminosity, which is one of the highest among
3CR quasars.  The radio and optical luminosities then follow the trend
found by Wills \& Brotherton (1995).

\subsubsection{Other Red, Scattered Light, Radio-Loud Quasars}

There are few quasars with unusual combinations of properties similar
to that of 3CR\,68.1.  As mentioned in \S\ 1, the quasar OI 287 
(Antonucci, Kinney, \& Hurt 1993; Goodrich \& Miller 1988; Rudy \& Schmidt 1988)
has a high, stable optical polarization, a quiescent optical flux,
blueshifted associated absorption, and is lobe dominant; OI 287 differs from
3CR\,68.1 in that its polarization position angle is parallel to its radio axis.
WN J0717+4611 (De Breuck et al. 1998) is another 
highly polarized red quasar, a steep spectrum radio source that is neither a
blazar nor radio-quiet QSO. Its 
polarization is perpendicular to the radio jet axis. 
It has been proposed that these quasars are partially obscured, with
red colors as a result of dust extinction, and polarization arising from
scattering along less obscured lines of sight.
3CR\,68.1 and objects such as these constitute a separate class of HPQ 
distinct from OVVs and radio-quiet QSOs.
There are other radio-loud type 2 (NLR dominated) objects known, of quasar-like
luminosities, with broad lines and continua seen in scattered light --
presumably `buried' quasars, e.g., Cygnus A (Ogle et al. 1997),
3C 324 (Cimatti et al. 1996), 3C 265 (Dey \& Spinrad 1996), 
and IRAS 09104+4109 (Hines \& Wills 1993).

\subsection{Geometry and Unified Schemes}

3CR\,68.1 is a classic quasar, judging by its luminosity, and broad UV emission
lines of typical equivalent width.
At 178 MHz it is one of the three most powerful 3CR quasars in the compilation
by Laing et al. (1983).
The extreme dominance of the powerful lobes in 3CR\,68.1, the largest projected
lobe separation of any 3CR source, the reddest continuum of any quasar, and the
highest scattered-light polarization of any 3CR source, immediately suggests a
connection with highly inclined AGN in Unified Schemes.  It is also of interest
for testing Unified Schemes as it is a member of the only sample with complete
optical identifications, selected by an essentially
orientation-independent quantity directly related to the power of the central
engine -- the low frequency (178 MHz) radio flux-density.

Barthel (1989) based the arguments for the unification of radio-loud quasars
and FR II radio galaxies on the 3CR sample, suggesting that radio galaxies are
quasars at axial inclinations greater than the half opening angle of a 
geometrically and optically thick dusty torus,
about 45$\arcdeg$, so that their quasar nuclei are obscured.
A variety of observational techniques have already shown that many 3CR 
radio galaxies harbor hidden quasar nuclei, and ionization and scattering 
`cones' have been found along the radio jet direction (e.g.,
Tran et al. 1998).
In principle the polarization and reddening properties of the 3CR sample could
be used to deduce the thickness of the obscuring torus as a function
of inclination.
Hill, Goodrich, \& DePoy (1996) partially perform this exercise by
looking for broad lines in the near infrared (where
dust reddening is less) for a complete sample of 11 3C radio sources
(8 narrow-line radio galaxies, two broad-line radio galaxies, plus 3C 273).
They find that the majority of objects have their broad-line regions reddened
by $A_V\approx3.0$, and their narrow-line regions by $A_V\approx1.5$.
We investigate the properties of 3CR\,68.1 in light of this Unification Scheme.

Figure 6 compares the observational results with the results of a
simple scattering model (after Brown \& McLean 1977, see also 
Wills et al. 1992) in which we assume a uniformly filled cone of 
wavelength-independent scatterers and a scattering optical depth of unity
appropriate for high polarization efficiency.
The geometry of the cone is defined by the inclination $i$ of its axis to the
observer's line-of-sight, and the cosine of its half-opening angle $\theta_c$.
The intrinsic polarization of light scattered towards the observer is $p_s$,
and Scattered/Direct is the observed ratio of scattered to direct light 
assuming no obscuration.  
Grid lines of various $i$ and $cos \theta_c$ are shown.
From the position of the models of Table 3 on the grid,
the cone half-opening angle is between $26\arcdeg$ and $46\arcdeg$, and the
inclination $i > 41 \arcdeg$.  For an ideal `virtual' cone with an axis
along the radio jets (the central engine axis) the
polarization position angle (E-vector) is perpendicular to the radio axis,
but the observed position angle for 3CR\,68.1 is $\approx60\arcdeg$ to the
radio axis.  This discrepancy suggests a
partially-filled virtual cone, or clumpy obscuration within a virtual cone.
In this case the cone determined from Figure 6 will be defined by the scatterers
that we `see'; it is tilted in the plane of the sky by 30$\arcdeg$, and
its opening angle is smaller than the size of the virtual cone.

The smaller inclinations consistent with the polarization of 3CR\,68.1 are
near the Barthel model dividing value of 45$\arcdeg$.  The modest reddening
towards the center (A$_V \approx 1$ -- 1.5) is also between values for radio
galaxies and UV-selected quasars.  
It may be helpful to realize that the observed continuum is especially
steep because the UV is shifted to observed wavelengths, not because
the reddening is especially high.  The line-of-sight would be
skimming the torus in this case.  However, the extreme lobe-dominance
and large lobe separation are not consistent with smaller inclinations.

The larger inclinations, corresponding to jets almost in the plane of the sky,
are consistent with the radio structure and the polarization results, but
in this case we would expect
the direct view of the center to be completely blocked by a thick torus.
One way out is for the torus to be geometrically thin, a possibility also
suggested for the other scattering-polarized radio-loud quasar OI\,287.
This might also tie in with the most luminous AGNs having the thinnest dusty
disks (e.g.,  Antonucci et al. 1993).

The strong, blueshifted, \ion{Mg}{2} absorption in 3CR\,68.1, 
identified by Aldcroft et al. (1993, 1994), and confirmed here, is
sufficiently deep that the absorbing gas must lie at least in the direct
path, and perhaps also in the scattered light path.  In Unified Schemes
lobe-dominant, high-inclination quasars are more likely to show absorption
along the line-of-sight (for 3C quasars -- Wills et al. 1995, see also
Aldcroft et al. 1993).  3CR\,68.1 follows this trend.

Several modifications of the simple Unified Model for radio galaxies 
and quasars, involving the effects
of absorption towards the central source, the radio core, the scatterers
and the NLR, are suggested in the case of 3CR\,68.1:
(i) even for highly inclined AGNs like the radio galaxies,
the radio core dominance is among the smallest (log\,R = $-3.2$).  
In the lower inclination case, a possibility is that
the reddened line-of-sight combined with the very weak radio core is the
result of radio free-free absorption by dusty, ionized gas.   This
could be tested by determining the shape of the radio spectrum, and
also by searching for X-ray absorption edges of O\,VII and O\,VIII
(as for IRAS\,13349+2438 -- Beichman et al. 1986; Brandt et al. 1997;
see also Leighly et al. 1997; Grupe et al. 1997).
(ii) the polarization position angle is not perpendicular to the radio
jets as discussed above, suggesting a partially filled or partially
obscured cone, and
(iii) the narrow-line (NLR) equivalent widths are similar to those in
our mean unreddened quasar spectrum, and also similar to the composite
lobe-dominated quasar
spectrum of Baker \& Hunstead (1995) formed from the Molonglo sample.
If 3CR\,68.1 were reddened by the amounts implied by our modeling, then
the continuum is suppressed by 20 to 30 times.  If the low-ionization NLR
extends well beyond the dusty torus, the NLR should suffer much less reddening
and EW(NLR) should be enhanced by 20 to 30 times.  The reddening of the
NLR in the Hill et al. radio-galaxies, the reddened scattered spectrum
of 3C\,68.1, and a partially obscured cone, all suggest significant
absorption.

We reach a limit to this simple interpretation if there is significant
forward-scattered, less-polarized light; in this case the obscuration we
derive from the steepening of the UV-optical spectrum would be an
underestimate.  But then the geometry of the simple Unified Scheme breaks
down.  In this regard, 3CR\,68.1 makes an especially interesting and 
important comparison with the (only) other scattered-light, luminous radio-loud
quasar OI\,287.  In OI\,287 the scattered light is thought to be scattered by 
electrons in the outer regions of a geometrically thin, dusty disk that 
obscures the center completely at UV-optical wavelengths.  This explains 
its polarization parallel to the radio jets.  3CR\,68.1 also shows 
polarization that is not exactly perpendicular to the radio jets;
a polarized component as in OI 287 provides an alternative explanation
to a partially filled or partially obscured scattering cone.

Overall 3CR\,68.1 fits well into the Unified Scheme, although it falls short
of being a textbook example in several potentially interesting respects.
These `shortcomings' may in the future provide opportunities to probe intrinsic
obscuring structures common to all quasars.

\section{Summary}

We have presented optical spectropolarimetry of 3CR\,68.1, a quasar with an 
unusual combination of extreme optical, radio, and polarization properties.
The polarization increases from $\approx$5\% at $\lambda_{rest}$ = 4000\AA,
to $>$10\% at $\lambda_{rest}$ = 2000\AA, with a nearly constant position 
angle of $\approx 53\arcdeg$.  Broad-emission lines are polarized
the same as the continuum.
We have measured parameters of the emission lines and absorption lines in the
total light spectrum. 

We have argued that scattering by dust or electrons is the polarization
mechanism, and that we see the quasar along two reddened lines of sight.
We have modeled this situation and compared our models with published 
photometry.  The implied reddenings are modest, $A_V = 0.7$ for the scattered
line of sight, and $A_V = 1.1$ to 1.6 for the direct line of sight, depending
on the intrinsic polarization of the scattered light and the possible 
contribution of light from an elliptical host galaxy implied by \ion{Ca}{2} 
K absorption.

We have interpreted the geometry in terms of unified models such that 
3CR\,68.1 is observed at a line of sight skimming the edge of an obscuring
torus.  While this interpretation appears correct and explains the
main properties 3CR\,68.1 very well, a few interesting issues remain
in the details of the geometry and in the placement and nature of both the 
scattering and absorbing material.

\acknowledgments

We thank Ron Quick, Randy Campbell, Tom Bida, and Bob Goodrich
for their assistance during our Keck run, Hien Tran for useful discussions, 
Mark Dickinson for providing some IRAF scripts, 
and Derek Wills for measuring the interstellar polarization toward 3CR\,68.1.  
We thank the editor and the referee, Dean Hines, whose suggestions improved
the paper. 
BJW gratefully acknowledges support through NASA LTSA grant number NAG5-3431.
The W. M. Keck Observatory is a scientific partnership between the
University of California and the California Institute of Technology,
made possible by the generous gift of the W. M. Keck Foundation.
This research has made use of the NASA/IPAC Extragalactic Database 
(NED) which is operated by the Jet Propulsion Laboratory, 
California Institute of Technology, under contract
with the National Aeronautics and Space Administration.
Work performed at the Lawrence Livermore National Laboratory is supported 
by the DOE under contract W7405-ENG-48.

\small
\voffset -0.5 truein
\def\s{$\,$}
\def\m{$-$}
\def\p{$+$}
\def\e{$\pm$}
\begin{deluxetable}{lccccc}
\tablewidth{42pc}
\tablenum{1}
\tablecaption{Emission-Line Measurements}
\tablehead{
\colhead{Line Identification} & \colhead{Center\tablenotemark{a}}& \colhead{Obsrved Flux} & \colhead{EW\tablenotemark{b}} & \colhead{FWHM\tablenotemark{c}} & \colhead{Galactic Factor\tablenotemark{d}} \\
\colhead{} & \colhead{(\AA)}& \colhead{($10^{-16} ergs\ s^{-1}\ cm^{-2}$)} & \colhead{(\AA)}& \colhead{($km\ s^{-1}$)} & \colhead{($A_V = 0.15$)}}
\startdata
Total \ion{C}{3}] $\lambda$1909 & 4251.0\e0.4 & 7.0\e1.0& 19.7\e1.8 & 1300\e200 & 1.45 \nl
Broad \ion{C}{3}] $\lambda$1909 & 4249\e3 & 5.9\e0.8 & 13\e1.8 & 7400\e800 & 1.45\nl
Narrow \ion{C}{3}] $\lambda$1909 & 4251.0\e0.4 & 3.0\e0.1 & 6.7\e2.2 & 800\e50 & 1.45\nl
\ion{C}{2}] $\lambda$2326 & 5183.5\e1.0 & 1.1\e0.4 & 1.1\e0.4 & 1300\e400 & 1.44 \nl
[\ion{Ne}{4}] $\lambda$2425 & 5399.5\e0.4 & 1.6\e0.1 & 1.3\e0.1 & 700\e50 & 1.39 \nl
Total \ion{Mg}{2} $\lambda$2800 \tablenotemark{e} & 6226\e2 & 40\e8 & 28\e5 & 4500\e500 & 1.30 \nl
Broad \ion{Mg}{2} $\lambda$2800 & 6227\e5 & 38\e5 & 25\e2 & 14400\e500 & 1.30 \nl
Narrow \ion{Mg}{2} $\lambda$2800 & 6226\e2 & 5.5\e0.5 & 3.1\e0.3 & 2000\e200 & 1.30\nl
\ion{He}{1} $\lambda$2830  & 6314.2\e0.8 & 1.4\e0.5 & 0.8\e0.3 & 1100\e200 & 1.30 \nl
\ion{O}{3} $\lambda$3133 & 6978\e2 & 1.2\e0.4 & 0.7\e0.2 & 1400\e200 & 1.27 \nl
\ion{He}{2} $\lambda$3203 (+ Fe II?) & 7139\e1.5 & 1.5\e0.3 & 7.9\e1.6 & 1600\e200 & 1.27 \nl
[\ion{Ne}{5}] $\lambda$3346 & 7452.6\e0.3 & 1.3\e0.4 & 0.8\e0.2 & 700\e100 & 1.25 \nl
[\ion{Ne}{5}] $\lambda$3426 + \ion{O}{3} $\lambda$3444\tablenotemark{f} & \nodata & \nodata & \nodata & \nodata & \nodata \nl
[\ion{O}{2}] $\lambda$3727 & 8309.0\e0.3 & 6.8\e0.4 & 3.5\e0.2 & 500\e40 &  1.24 \nl
[\ion{Ne}{3}] $\lambda$3869 + H8 + He I $\lambda$3868 & 8623.3\e0.3 & 6.8\e1.0 & 3.5\e0.6 & 600\e70 & 1.23 \nl
[\ion{Ne}{3}] $\lambda$3967 + H$\epsilon$\tablenotemark{g} & \nodata & \nodata & \nodata & \nodata & \nodata \nl

\enddata

\tablenotetext{a}{Air wavelengths.}

\tablenotetext{b}{Corrected to rest-frame.}

\tablenotetext{c}{Corrected for the instrumental resolution of $\sim$10\AA.}

\tablenotetext{d}{Flux densities should be multiplied by this factor 
to correct for Galactic reddening.}

\tablenotetext{e}{The absorption lines and \ion{He}{1} $\lambda$2830 were 
interpolated across in order to estimate the integrated flux and equivalent
width for \ion{Mg}{2}.  The emission line peak was estimated based on the
unabsorbed narrow component wings.}

\tablenotetext{f}{The [\ion{Ne}{5}] $\lambda$3426 + \ion{O}{3} $\lambda$3444
blend, often easily measured in good spectra of quasars and radio galaxies,
falls at the same wavelength as atmospheric A band absorption (which
we were unable to remove well with our standard star).}

\tablenotetext{g}{The [\ion{Ne}{3}] $\lambda$3967 + H$\epsilon$ blend is
coincident with \ion{Ca}{2} H absorption, and also with a
strong night sky feature.  This blend cannot be reliably measured.}

\end{deluxetable}
\normalsize

\begin{deluxetable}{lcccc}
\tablewidth{280pt}
\tablenum{2}
\tablecaption{Absorption-Line Measurements\tablenotemark{a}}
\tablehead{
\colhead{Identification} & \colhead{Center\tablenotemark{b}} & \colhead{$\lambda_{rest}$} & \colhead{z$_{abs}$} & \colhead{EW\tablenotemark{c}} \\
\colhead{} & \colhead{(\AA)}& \colhead{(\AA)}&\colhead{} & \colhead{(\AA)}}
\startdata
Fe II $\lambda$2382 & 5301.7\e0.7 & 2382.0 & 1.2257(3) & 1.0\e0.3 \nl
Fe II $\lambda$2586 & 5757.3\e0.1 & 2585.9 & 1.2264(1) &  1.0\e0.2 \nl
Fe II $\lambda$2600 & 5785.5\e0.1 & 2599.4 & 1.2257(1) & 2.0\e0.2 \nl
Mg II $\lambda$2796 & 6221.0\e0.5& 2795.5 & 1.2254(2) &  6.5\e0.2 \nl
Mg II $\lambda$2803 & 6238.3\e0.2& 2802.7 & 1.2258(1) &  4.4\e0.2 \nl
Mg I $\lambda$2853 & 6348.3\e0.6 & 2852.1 & 1.2258(2) & 1.1\e0.2 \nl
He I $\lambda$3187.7?& 7093.0\e1.0 & 3187.7 & 1.2251(3) & 1.5\e0.4 \nl
\ion{Ca}{2} K & 8767.0\e0.5 & 3933.7 & 1.2287(1) & 5.0\e0.8  \nl
\enddata
\tablenotetext{a}{Uncertanties quoted denote the range of acceptable 
measurements rather than formal 1 $\sigma$\ fitting errors.}
\tablenotetext{b}{Air wavelengths.}
\tablenotetext{c}{Equivalent widths (EWs) are in the observed frame.}  
\end{deluxetable}

\begin{deluxetable}{lcccc}
\tablewidth{0pt}
\tablenum{3}
\tablecaption{Model Parameters}
\tablehead{
\colhead{Model} & \colhead{P$_s$} & \colhead{Galaxy} & \colhead{Direct $A_V$} & \colhead{Scattered/Direct} 
}
\startdata
A & 80\% & No & 1.2 & 0.04\nl
B & 50\% & No & 1.3 & 0.06 \nl
C & 20\% & No & 1.6 & 0.13 \nl
A$_G$ & 80\% & Yes & 1.1 & 0.06 \nl
B$_G$ & 50\% & Yes & 1.2 & 0.09 \nl 
C$_G$ & 20\% & Yes & 1.5 & 0.17 \nl
\enddata
\end{deluxetable}

\newpage
\psfig{file=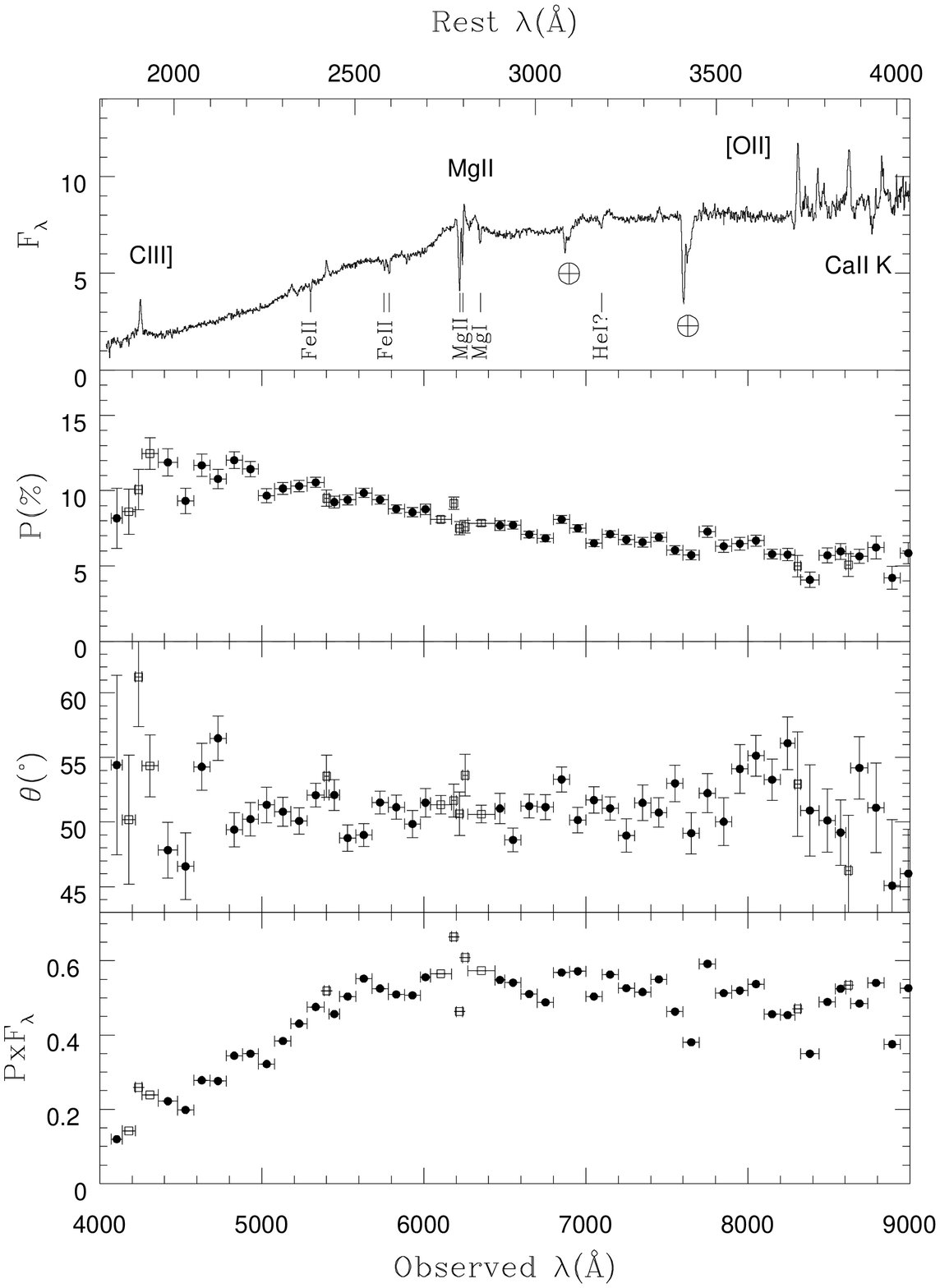,height=17.5cm}
\figcaption{Spectropolarimetric results for 3CR\,68.1. The top abscissa
shows rest-frame wavelengths, while the bottom abscissa shows observed-frame
wavelengths, both in \AA.  The top panel is the total flux spectrum (in
10$^{-17}$ ergs s$^{-1}$ cm$^{-2}$ \AA$^{-1}$), with some prominent features
labeled and narrow absorption lines marked from below.
The second panel from the top shows the percentage polarization.
The third panel is the polarization position angle in degrees.
The bottom panel shows the polarized flux, the product of the binned 
fractional polarization and binned total light spectrum.
Vertical error bars are 1 $\sigma$, while the horizontal bars show the extent
of the bins as described in the text.}

\newpage  
\psfig{file=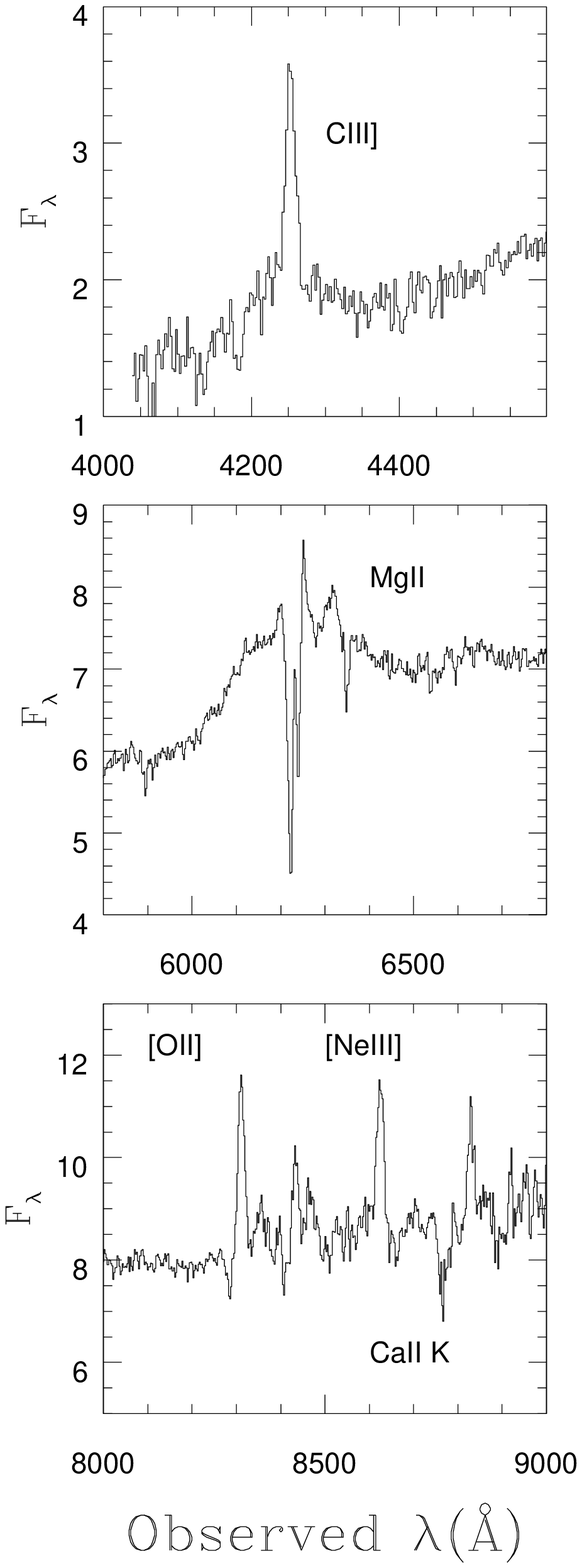,height=20cm}
\figcaption{Detail of several regions of the total light spectrum.
Wavelengths are observed frame, while flux units are $10^{-17}$
ergs s$^{-1}$ cm$^{-2}$ \AA$^{-1}$.}

\newpage
\psfig{file=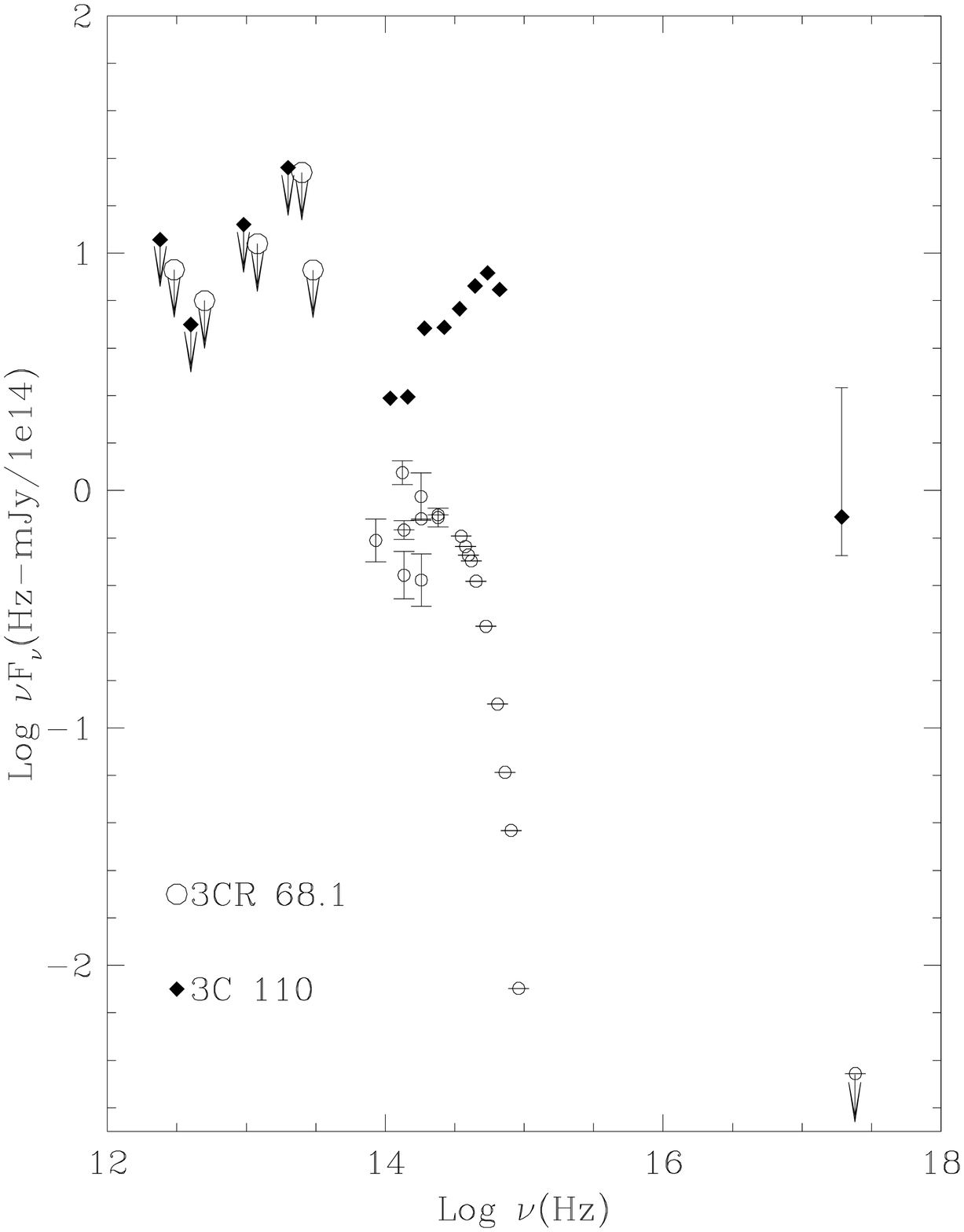,height=17cm}
\figcaption{The (observed-frame) spectral energy distribution for 3CR\,68.1.
The optical points are from Neugebauer et al. (1979) and
Smith \& Spinrad (1980).  Mid and near-IR points are from
Neugebauer et al. (1979), Stein \& Sitko (1984), and  Rieke et al. (1982).
The X-ray upper limit comes from the RASS survey limit, after checking the
count rates and standard deviations in the 3CR\,68.1 region of sky, and using
standard flux-density conversions, assuming F$_{\nu} \propto
{\nu}^{-2.25}$\ and a Galactic hydrogen column density
N$_H = 6 \times 10^{20}$\,cm$^{-2}$\ (Voges et al. 1996).
The data for 3C 110, shifted to the observed frame
of 3CR\,68.1, comes from the Atlas of Elvis et al. (1994),
with the near-IR and optical data for only the October 1984 epoch plotted.}

\newpage
\psfig{file=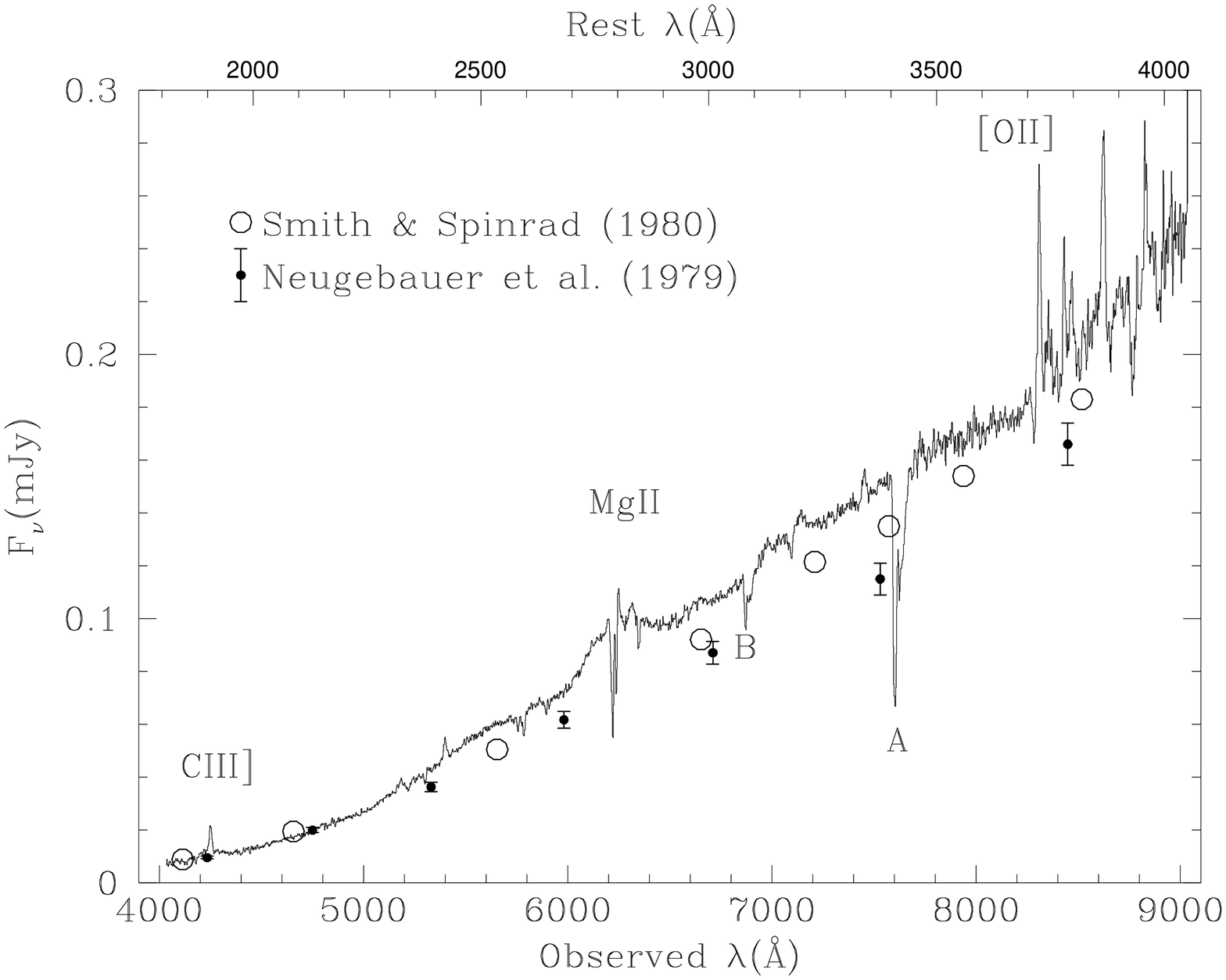,height=18cm,width=18cm,angle=-90}
\figcaption{Comparison our total flux density spectrum with those of
Neugebauer et al. 1979 ({\em dark circles}, error bars as
given by Neugebauer et al. 1979) and Smith \& Spinrad 1980
({\em open circles}).}

\newpage
\psfig{file=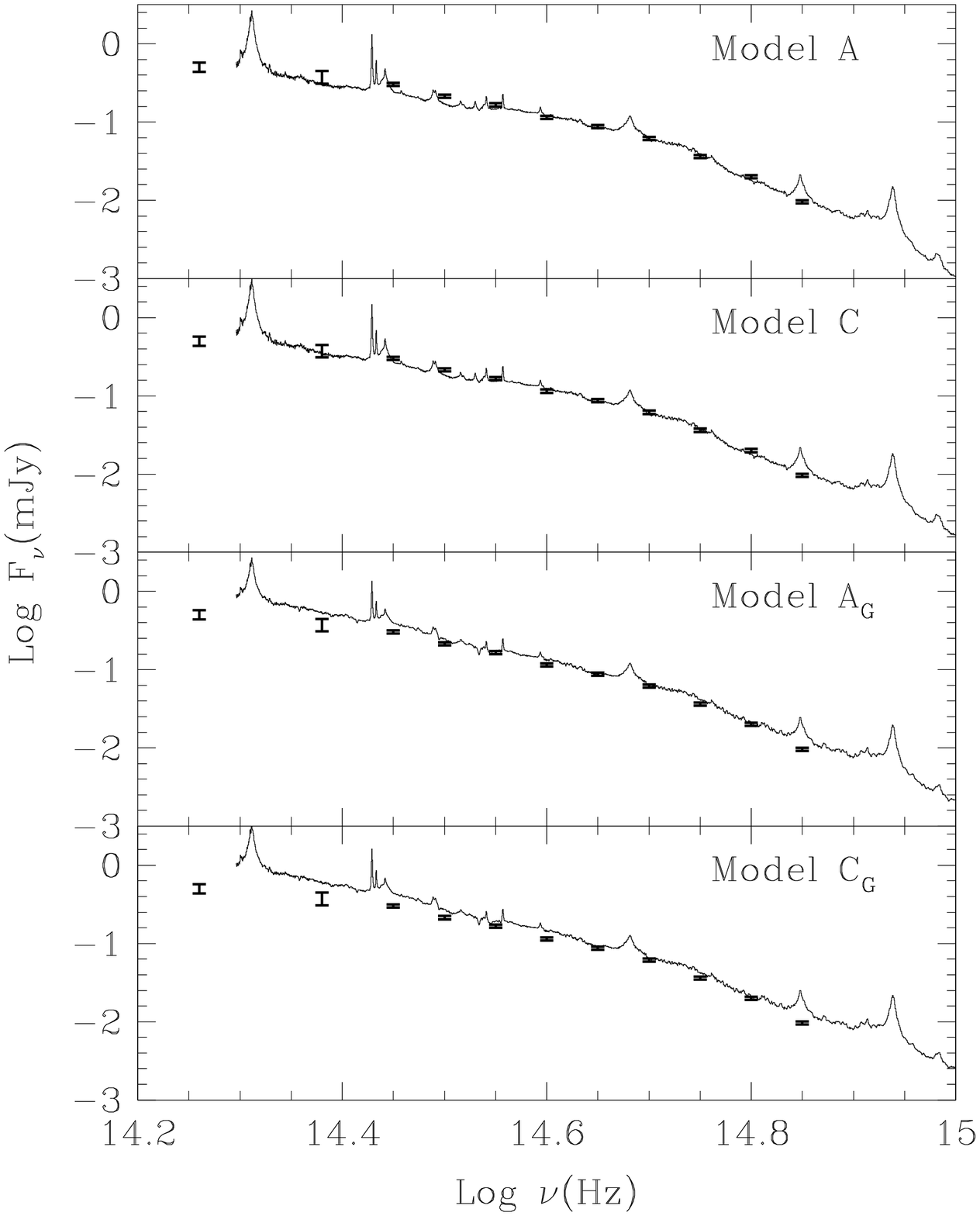,height=18cm}
\figcaption{The extreme models from Table 3 and the text, with and without
a host galaxy component, compared with the Neugebauer et al. (1979) 
3CR\,68.1 photometry, $B$ to $H$.  The bold errors bars are as given by 
Neugebauer et al.}

\newpage
\psfig{file=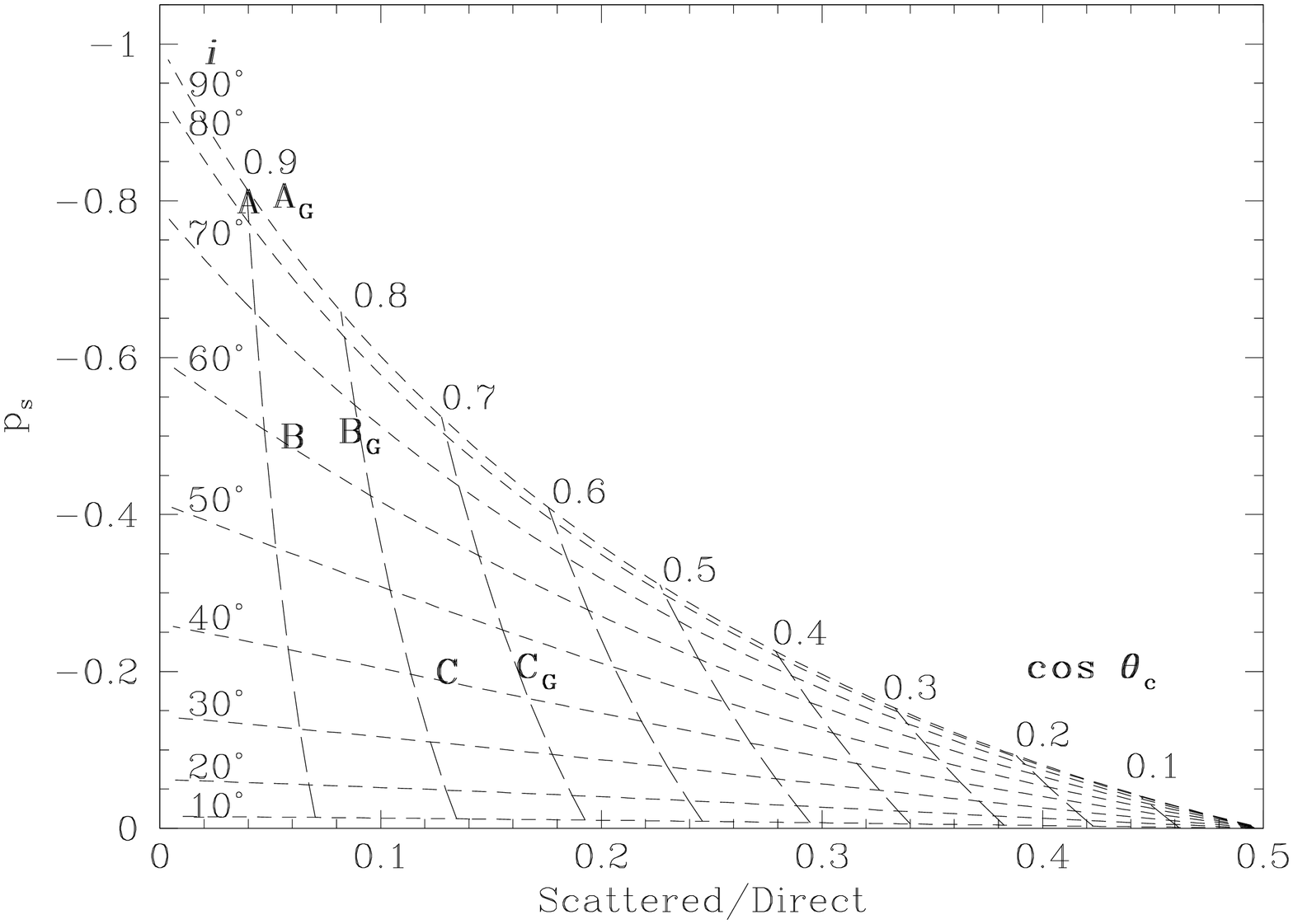,height=17cm,width=18cm,angle=-90}
\figcaption{A simple scattering model.  The fractional linear polarization 
of scattered light ($p_s$) is plotted against the ratio of scattered to 
direct light for a uniformly filled cone of wavelength-independent scatters.
The model includes no reddening or obscuration.
The negative sign on $p_s$ indicates that the electric vector of the 
polarized light is perpendicular to the cone axis.
Model lines show the effects of varying $i$, the inclination
of the axis to the line of sight, and the cosine of the 
half opening angle of the cone, $\theta_c$.  Models A, B, and C correspond to 
our data on 3CR\,68.1 under the assumptions that the host galaxy light
is negligible and $p_s$ = 80\%, 50\%, and 20\%, respectively.
The $G$\ subscripts are for models with a host galaxy contribution.}

\end{document}